\documentstyle[aps,twocolumn]{revtex}
\input psfig

\begin{document}

\draft

\title{Measurement of Arbitrary Observables of a Trapped Ion}

\author{S.~A.~Gardiner, J.~I.~Cirac$^{*}$, and P.~Zoller }
\address{Institut f{\"u}r Theoretische Physik,
Universit{\"a}t Innsbruck, 6020
Innsbruck, Austria}
\address{*Departomento de Fisica, Universidad da Castilla-La Mancha,
13071
Ciudad Real, Spain}
\date{\today}
\maketitle

\begin{abstract}

We describe a method to perform a single quantum measurement of an 
arbitrary observable of a single ion moving in a harmonic potential.
We illustrate the measurement procedure with explicit examples, namely
the
position and phase observables.
\end{abstract}


According to Quantum Mechanics every observable ${\cal A}$ is
represented by
a Hermitian operator $\hat{A}$, and the eigenvalues $a_k$ of this
operator
represent possible outcomes of a single measurement of ${\cal A}$
\cite{galindo}.  
Given a
state $|\phi \rangle $ of our system the probability for obtaining
$a_k$ 
is $P_k=|\langle \psi _k|\phi \rangle |^2$ where $|\psi_k\rangle $ is
the eigenvector associated with $a_k$, $\hat{A}|\psi_k\rangle
=a_k|\psi _k\rangle$ \cite{degenerate}. While Quantum Mechanics states
that {\em any\/} observable can be measured in principle, in practice
for a
given system only a few observables like momenta, energy etc.\ are
accessible in the laboratory.  
In the present letter we will show how in a
model system of an ion moving in a harmonic trap the measurement of an
{\it
arbitrary\/} observable of the ion motion can be implemented.
In this way one can perform {\em single measurements\/} of observables
thus far 
considered inaccessible, such as the Pegg-Barnett phase operator
\cite{pegg}, or any 
combination of position and momentum, such as angular momentum etc. 
We emphasize that 
in contrast to quantum tomography \cite{vogeltomography}, 
where the full density matrix of a system is determined by {\em
repeated\/}
preparations and measurements, we are considering a {\em single\/}
measurement
of an arbitrary observable.

Trapped ions are exceptionally well 
suited systems to study fundamental aspects
of quantum mechanics \cite{review}.
 By coupling laser light to the internal degrees of
freedom, a single ion can be cooled to the vibrational ground state of
the
trap, and its motion can be manipulated coherently to generate
nonclassical
states of motion. State measurement of the internal degrees of freedom
can
be carried out with essentially $100\%$ efficiency using quantum jump
techniques\cite{review}. These unique properties
have stimulated a series of fundamental experiments from generation of
nonclassical states of motion \cite{revivals} and Schr\"{o}dinger cat
states
\cite{cats}, to
implementation of quantum gates \cite{gates}. In addition, there have
been proposals for
tomographic measurements of the atomic motional density matrix
\cite{tomography}, and quantum
reservoir engineering \cite{reservoir}. 

To carry out the single measurement of an arbitrary observable ${\cal
A}$ we
will employ the following tools which, as will be shown below, are
readily
implemented in an ion trap: first, the ability to synthesize {\it any}
motional state $|\psi \rangle $ of the ion starting from the vibrational
ground state $|0\rangle$, {\it i.e.} $|\psi \rangle
=\hat{U}_{\psi}|0\rangle $
where $\hat{U}_{\psi}$ is a unitary operator; second, the ability to 
perform a filtering measurement \cite{galindo} to distinguish whether
the
ion is in the vibrational ground state or not. 

To illustrate the procedure we assume the ion to be in the unknown state
$| \phi \rangle $ \cite{mixed,rwa}.  Let us denote by $\hat{U}_k$ the
unitary
time evolution operator which generates the eigenstate $|\psi _k\rangle
=
\hat{U}_k|0\rangle$ of the observable that we want to measure. We first
transform the state of the ion by the inverse
transformation to $|\chi _0\rangle =\hat{U}_0^{\dagger }|\phi \rangle $,
and
measure whether the ion is in the ground motional state or not, after
which
$\hat{U}_{0}$ is applied. 
If the ion was found to be in the ground motional state, the state after
this
procedure will be $|\psi_{0}\rangle$, which implies that we have
measured
$a_{0}$.
The
probability for this to happen is
\begin{equation}
|\langle 0|\chi _0\rangle |^2=|\langle \psi _0|\phi \rangle |^2\equiv
P_0.
\end{equation}
which agrees with the probability $P_0$ of measuring $a_0$. In the case
that
we do not measure the ion in $|0\rangle$, its state is projected onto
\begin{equation}
|\phi_{0}\rangle =
\left( 1 - |\psi_{0}\rangle \langle \psi_{0}|\right)|\phi \rangle/
|\!|\ldots|\!|,
\end{equation} 
where $|\!|\ldots|\!|$ serves to normalize the state.
We then transform the state of the ion according to
$|\chi _1\rangle =\hat{U}_1^{\dagger }|\phi_0 \rangle $,
measure whether it is in the ground motional state or not, and  apply
the
unitary evolution operator $\hat{U}_{1}$.
If the ion is found in $|0\rangle$, the state after this step will be
$|\psi_{1}\rangle$.
The corresponding probability will be equal to 
\begin{equation}
(1-P_0) |\langle 0|\chi_1\rangle|^2 = |\langle \psi_1|\phi\rangle|^2
\equiv P_1,
\end{equation}
{\it i.e.} it coincides with the probability of measuring $a_1$. In the
case
that
we do not measure the ion in $|0\rangle$, we apply the unitary
evolution operator $\hat{U}_{2}^{\dagger}$, and continue
in the same vein. After $k$ steps, the probability of measuring 
the ion in its ground state will be $P_k$, and the state of the ion will
be
projected onto the state $|\psi_{k}\rangle$. In this way, we will
measure
the ion in one of the possible eigenstates $|\psi_k\rangle$, which
is equivalent to obtaining as a result the value $a_k$ of the
observable ${\cal A}$. In practise this procedure must be carried out in
a
finite number of steps, so we restrict ourselves to a finite dimensional
Hilbert space.

Let us now show how the above procedure can be implemented in the
case of an ion trapped in a harmonic potential. For the sake of 
simplicity, we will consider the one--dimensional case; the 
generalization to more dimensions will be described elsewhere.
Here we will discuss how to prepare arbitrary states of the
ion motion out of the ground state $|0\rangle$, i.e. how to implement
and derive the unitary evolution operators $\hat U$ using
laser pulses. This method is  
conceptually similar to the one proposed by Law and Eberly in the
context of cavity QED \cite{eberly}. We will also simulate numerically
the measurement 
procedure described above for the case of a trapped ion. 

We consider a single two--level ion 
of ground level $|g\rangle$ and excited level $|e\rangle$, where
spontaneous
emission is negligible \cite{review},
trapped in a harmonic potential and 
interacting with laser fields. 
For a travelling wave configuration the Hamiltonian of this
system in 
the interaction picture is:
\begin{equation}
\label{Hamilton1}
\tilde{H}_{\mbox{\scriptsize \it tr}} =
\frac{\Omega}{2}\left[\sigma^{+}e^{-ik\hat{x}(t) + i\phi
-i\Delta t} + \mbox{H.c.}\right],
\end{equation}
where
$k\hat{x}(t)  = \eta\left(\hat{a}e^{-i\nu t} + \hat{a}^{\dagger}e^{i\nu
t}\right)$, $\Delta = \omega_{L} - \omega_{0}$ is the detuning, 
$\omega_{L}$ is
the laser frequency, $\omega_{0}$ is the frequency gap between the
ground and excited atomic levels, $\nu$ is the trap
frequency, $\eta$ is the Lamb-Dicke parameter, $\Omega$  is the laser
Rabi frequency, and $\hat{a}^{\dagger}$ and
$\hat{a}$ are respectively the phonon creation and annihilation
operators.
 
The method for synthesizing states consists of starting out with a
cooled ion in state $|g,0\rangle$, and by an appropriate sequence of
laser pulses
to coherently distribute this amplitude to formed a desired arbitrary
superposition: $\hat{U}|g,0\rangle = \sum_{n=0}^{N}c_{n}|g,n\rangle$ 
\cite{rwa}, where we are restricted to ${\cal H}_{N+1}$.
The {\em calculational\/} method assumes we are carrying out the
opposite; we begin with a given arbitrary superposition and coherently
coalesce it
into the state $|g,0\rangle$. This is carried out as follows: given that
the
highest value of $n$ is $N$, we use a laser pulse to push all of the
population
of $|g,n\rangle$ into $|e,n-1\rangle$ (a ``diagonal'' pulse), and then
all
of the population of $|e,n-1\rangle$ into $|g,n-1\rangle$ (a
``vertical''
pulse), and so on, keeping careful track of what is happening to all of
the
other populated levels of the ion all the while. Inverting the unitary
operations describing these pulses and applying them in reverse order
will give us
$\hat{U}$, as desired.

For the ``vertical'' pulse, we tune the laser on resonance ($\Delta =
0$). When $\Omega$ is sufficiently small we can assume that there are no
off-resonant transitions \cite{review}, that is to say the level
structure of the ion can be
considered to be a series of parallel, isolated, two-level systems.
Precisely what is
meant by ``sufficiently small'' will be detailed later. Assuming this,
and
transforming to a rotating frame ($\hat{U}_{\mbox{\scriptsize rfv}} =
e^{i\nu\hat{a}^{\dagger}\hat{a}t}$) one can easily show that the
Hamiltonian
(\ref{Hamilton1}) reduces to:
\begin{equation}
\tilde{H}_{\mbox{\scriptsize \it tr}}
\approx
\sum_{n=0}^{\infty}\frac{1}{2}\left(\Omega_{n}\sigma^{+}e^{-i\phi}
+ \Omega_{n}^{*}\sigma^{-}e^{i\phi}\right)|n\rangle\langle n|,
\label{tvert}
\end{equation}
where 
$\Omega_{n} = \Omega\langle n| e^{-i\eta\left(\hat{a} +
\hat{a}^{\dagger}\right)}|n\rangle$ is the effective Rabi frequency,
given by:
\begin{equation}
\Omega_{n} = 
\Omega\sum_{k = 0}^{n}
\left(
\begin{array}{c}
n \\
k
\end{array}
\right)
\frac{\left(-\eta^{2}\right)^{k}e^{-|\eta|^{2}/2}}{k!}.
\label{omegavert}
\end{equation}

The general unitary operation describing transfer of population between
a ground and an excited atomic state (essentially a rotation) can be
described by two parameters, $\theta$ and $\chi$, defined by:
\begin{equation}
\hat{U} =
\left(
\begin{array}{cc}
\cos\theta & -e^{i\chi}\sin\theta \\
e^{-i\chi}\sin\theta & \cos\theta
\end{array}
\right),
\label{unitary}
\end{equation}
where $|g\rangle = (0,1)$, $|e\rangle = (1,0)$. For
$\hat{U}(r_{g}e^{i\psi_{g}}|g\rangle + r_{e}e^{i\psi_{e}}|e\rangle) =
\sqrt{r_{g}^{2}+ 
r_{e}^{2}}e^{i\psi_{f}}|g\rangle$ (the desired operation in this case),
$\tan\theta = r_{e}/r_{g}$, and $\chi = \psi_{e} - \psi_{g}$.
$\psi_{f}$ is the phase of the amplitude of the {\it finally\/}
populated
state.

From (\ref{unitary}) and (\ref{tvert}) we can deduce that
$\Omega_{n}t/2 = \theta \; \forall \; n \geq 0$.
Thus, if we calculate $\theta $ to clear the ground state of the ``top''
two level system (that system still populated having the largest number
$N$ of phonons) $\theta_{N}$, for those systems possessing fewer ($n$)
phonons,
$\theta_{n} = \Omega_{n}\theta_{N}/\Omega_{N}$. For each two-level
system $\phi = \chi +\pi/2$.

In the case of the ``diagonal'' pulse, we set $\Delta = -\nu$ (tune the
laser
to the lower sideband) and transform to a
different rotating frame ($\hat{U}_{\mbox{\scriptsize  rfd}} =
e^{i\nu\left(\hat{a}^{\dagger}\hat{a} + \sigma_{z}/2\right)t}$).
In this case, the Hamiltonian (\ref{Hamilton1}) can be approximated by
\begin{equation}
\tilde{H}_{\mbox{\scriptsize \it tr}} \approx
\sum_{n=0}^{\infty}\frac{\Omega_{n}'}{2}\sigma^{+}e^{-i\phi}|n\rangle\langle
n+1| + \mbox{H.c.},
\label{diag}
\end{equation}
where
$\Omega_{n}'=\Omega\langle n|e^{-i\eta\left(\hat{a} +\hat{a}^{\dagger}
\right)}|n+1\rangle$ is the effective Rabi frequency, given by:
\begin{equation}
\Omega_{n}' = -i\Omega\sum_{k=0}^{n}
\left(
\begin{array}{c}
n+1 \\
k+1
\end{array}
\right)
\frac{(-1)^{k}\eta^{2k+1} e^{-|\eta|^{2}/2}}{\sqrt{n+1}k!}.
\label{omegadi}
\end{equation}
Note that the effective Rabi frequencies $\Omega_{n}$ and 
$\Omega_{n}'$ are calculated from the full expansion
of the exponential $e^{-ik\hat{x}t}$, and that our proposed scheme is
therefore
not restricted to the situation where $\eta \ll 1$ (Lamb-Dicke limit),
in contrast to previous work \cite{review} and cavity--QED
\cite{eberly}.

Referring again to (\ref{unitary}), the desired operation is
$\hat{U}(r_{g}e^{i\psi_{g}}|g\rangle +
r_{e}e^{i\psi_{e}}|e\rangle) = \sqrt{r_{g}^{2}+ 
r_{e}^{2}}e^{i\psi_{f}}|e\rangle$, so that
$\tan\theta = r_{g}/r_{e}$, $\chi = \psi_{e} - \psi_{g} + \pi$. We thus
deduce:
$\Omega_{n}'t/2 = \theta \; \forall \; n \geq 0$;
$\theta_{n}$ (applies to the system of $|g, n+1\rangle$ and
$|e,n\rangle$)
$ = \Omega_{n}'\theta_{N}/\Omega_{N}'$; $\phi = \chi$.
To invert (\ref{unitary}) for both pulses
we simply add $\pi$ to       the calculated $\chi$.

Let us consider when
the use of the approximate Hamiltonians which express the ion as a
simple sum of non-interacting two-level systems is justified. 
For the sake of simplicity, we will give an estimate assuming the
Lamb-Dicke
limit, defined as $\eta\ll 1$. In this case the condition 
$
t_{\mbox{\scriptsize total}}\Delta E \ll 1
$
must be fulfilled, where
$\Delta E$ is the level shift induced by the next most important (off
resonant)
pulse when the laser is applied. Using second-order perturbation theory,
this
can be calculated to be $V^{2}/\Delta$, where $V$ is the effective Rabi
frequency of this secondary pulse and $\Delta$ is the detuning from the
frequency gap of the two levels it couples together. 
The
time taken for each ``vertical'' pulse $t= 2\theta/\Omega$, and for each
``diagonal'' pulse $t = 2\theta/(\eta\sqrt{n}\Omega)$, where $n$ is the
number of phonons in the ground state of the given two--level system.
Taking the variable $\theta$ to be of order 1, we need
$
\Omega/2 \ll \nu/mn\eta^2
$
for each ``vertical'' two level system, and
$
\Omega/2 \ll \nu \eta\sqrt{n}/m
$
for each ``diagonal'' two level system, where $m$ is the number of
pulses applied to that given system while it is occupied. To make single
inequalities for each two-level system, we take the maxima of $mn =
[(N+1)/2]^2$ and
$m/\sqrt{n} = N$, where $N$ is the maximum value of $n$ in the state to
be synthesized. 

We now define a ``quality factor'' $q$, 
and use this to calculate the laser Rabi frequencies
$\Omega/2 = q4\nu/[(N+1)\eta]^2$ and $\Omega/2 = q\nu\eta/N$ for the
cases of
the ``vertical'' and ``diagonal'' pulses respectively. These expressions
replace the original inequalities described above, with $q$ determining
the ``quality'' of the laser pulse ({\it i.e.} the smaller the value of $q$,
the greater the validity of the approximate Hamiltonians (\ref{tvert})
and 
(\ref{diag}), as appropriate). 

\begin{figure}[htbp]
\begin{center}\
\psfig{file=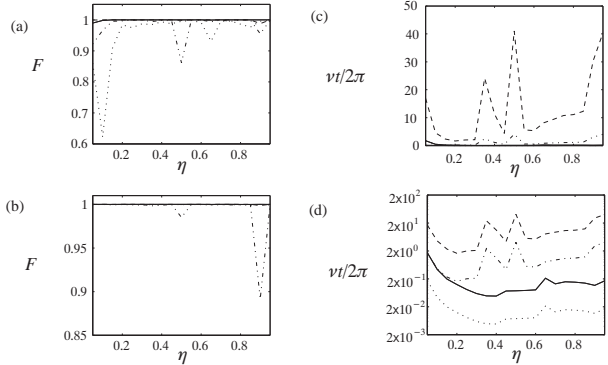,width=80mm}
\end{center}
\caption{For all four plots: 
solid line, phase state with $N=8$ and $\phi =2$ radians, 
synthesized when $q=0.01$; 
dotted line, 
same state 
synthesized when $q=0.1$; 
dashed-dotted, phase state with $N=32$ and $\phi=2$, 
synthesized when $q=0.01$;
dashed line, 
same state 
synthesized when $q=0.1$.
(a) Fidelity of synthesis of the above states 
using a travelling wave laser configuration for a range of $\eta$.
(b) Fidelity of synthesis using a standing wave configuration.
(c) Magnitude of the dimensionless quantity $\nu t/2\pi$ required for
the 
synthesis
of the states over a range of $\eta$ (identical for
both lasere configurations).
 (d) Same as (c), except using a logarithmic scale.}
\label{olap}
\end{figure}
The plots in Fig.\ \ref{olap} show the results of simulations creating
Pegg-Barnett phase states 
$|\phi_{k}\rangle = \sum_{n=0}^{N} e^{i\phi_{k}n}|n\rangle/\sqrt{N+1}$
(where $\phi_{k} = 2\pi k/(N+1)$) \cite{pegg}. Below we will measure the
Pegg-Barnett phase operator:
\begin{equation} 
\hat{\phi} = \sum_{k=0}^{N} \phi_{k}|\phi_{k}\rangle\langle\phi_{k}|,
\label{phaseop}
\end{equation}
and
will therefore obviously need to know how to synthesize its eigenstates
(the
phase states described above).
For these simulations we have used the exact Hamiltonian
(\ref{Hamilton1})
for various values of $q$ and $\eta$. 
We define the {\it fidelity} $F$ of the synthesis by
$F = |\langle \mbox{desired
state}|\mbox{synthesized state}\rangle|^2$, as a measure of how close
the
synthesized state is to what it is supposed to be. 
Note that our calculations as to the circumstances when the use of the
approximate Hamiltonians (\ref{tvert},\ref{diag}) is justified is
restricted to
the Lamb-Dicke limit, including our use of the quality factor $q$. We
nevertheless continue to use $q$ to define the laser Rabi frequencies up
to
$\eta = 0.95$ in Fig.\ \ref{olap}, for the sake of comparison.

As indicated in Fig. \ref{olap}, a completely analogous derivation can
be
carried out for a standing wave laser configuration \cite{review}. 
A ``vertical'' pulse is achieved when
$\Delta = 0$ and the ion is at an antinode. A  ``diagonal''
pulse is achieved when $\Delta = -\nu$ and the ion is at a node. The
approximate Hamiltonians are completely identical to the travelling wave
case
(\ref{tvert},\ref{diag}), except that
$e^{-i\eta(\hat{a}+\hat{a}^{\dagger})}$ is
replaced by $\cos(\eta[\hat{a}+\hat{a}^{\dagger} ])$ and
$\sin(\eta[\hat{a}+\hat{a}^{\dagger}])$ for the
``vertical'' and ``diagonal'' pulses respectively.
Thus, the constraints are somewhat different.
In the case of the ``vertical'' pulses, when the cosine function
is expanded to a power series
there are obviously only even powers of $\eta(\hat{a} +
\hat{a}^{\dagger})$.
This means that when the ``vertical'' pulse is the pulse desired, the
secondary
pulse is not the ``diagonal'' pulse (which does not exist) but the pulse
coupling $|g, n\rangle$ to $|e,n-2\rangle$, which is much smaller.
Similarly in
the case of the ``diagonal'' pulse, (sine expansion, and thus no odd
powers of
$\eta(\hat{a} + \hat{a}^{\dagger})$) the secondary pulse is that
coupling
$|g,n\rangle$ to $|e,n+1\rangle$, which is also small. Thus the
constraints on
the intensity of the laser are much less restrictive. We nevertheless
use the
travelling wave constraints for constructing the states shown in Fig.\
\ref{olap}, to show that significantly better results can be achieved
using a
standing wave configuration, where the circumstances are otherwise
identical.
Note also that we are generally used to thinking of the regime where the
Lamb-Dicke limit can be assumed \cite{review} as being universally
optimal.
If we look at the plots in Fig.\ \ref{olap} however, 
we can see that higher values of
$F$ and lower values of $\nu t$ are achieved for comparitively large
$\eta$.

\begin{figure}[htbp]
\begin{center}\
\psfig{file=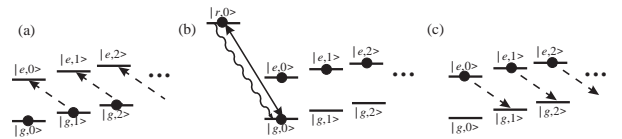,width=80mm}
\end{center}
\caption{The filtering measurement process. (a) The amplitudes of all
the
ground states $|g,n+1\rangle$ except for $|g,0\rangle$ are adiabatically
transferred to the $|e,n\rangle$ states. (b) 
The laser is tuned to the decaying $|g\rangle \leftrightarrow |r\rangle$
transition.
The presence of fluorescence will indicate that the ion was in the state
$|g,0\rangle$ {\protect \cite{review}}. 
(c) If no photon
is spontaneously emitted, then the original amplitude of the state
$|g,0\rangle$ is projected out, and the population of the excited states
$|e,n\rangle$
is adiabatically transferred back to the states $|g,n+1\rangle$.}
\label{fluor}
\end{figure}
The other tool we require in our measuring procedure is the ability to
perform
filtering measurements. First we shift all of the population in states
$|g,n+1\rangle$ to the states $|e,n\rangle$ by adiabatic passage 
\cite{adiabfock} (where $n \geq 0$) so that all amplitudes with the
exception
of that for $|g,0\rangle$ are transferred unchanged (see Fig.\
\ref{fluor}). Next we use
quantum jump techniques \cite{review} to determine whether $|g,0\rangle$
is populated \cite{cool}. Finally, if we do not detect population in
state 
$|g,0\rangle$, we
use adiabatic passage to restore the amplitudes in states $|e,n\rangle$
back 
to the states $|g,n+1\rangle$. Adiabatic passage is required so that the
population can be transferred independent of the Rabi frequencies.

\begin{figure}[htbp]
\begin{center}\
\psfig{file=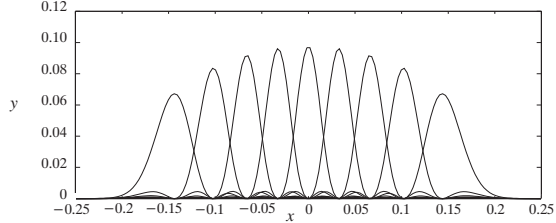,width=80mm}
\end{center}
\caption{Plot showing the spread of the discrete position eigenstates
$|x_{k}\rangle$ of 
${\cal H}_{9}$.
$x$ is in wavelengths$/\eta$, and 
$y =|\langle x|x_{k}\rangle|^{2}$. $|x_{0}\rangle$ is the leftmost peak,
and
$|x_{9}\rangle$ the rightmost.}
\label{poseig}
\end{figure}
\begin{figure}[htbp]
\begin{center}\
\psfig{file=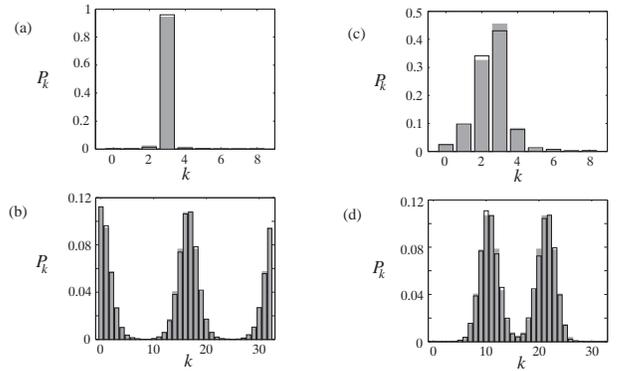,width=80mm}
\end{center}
\caption{
Measurement of the the phase (a,b) and position (c,d) observables
{\protect \cite{rwa}}.
(a,c) The ion is initially in the phase state with $N = 8$ and $\phi =
2$
radians; for these simulations $q = 0.1$  and $\eta = 0.5$. 
(b,d) The ion is initially  in the Schr\"{o}dinger Cat state with $N=32$
and
$\alpha = 1.5$; for these simulations $q=0.1$ and $\eta= 0.2$.
For each plot,
black outline bars represent the results of simulation of the full
Hamiltonian,
as given by Eq.\ \ref{Hamilton1} and grey solid bars represent ideal 
probabilities.}
\label{cat}
\end{figure}

We have simulated our measurement process, where one attempts to
determine the 
phase and
position observables of particular states. To this end we used as bases:
the
eigenstates 
of the Pegg-Barnett phase operator (\ref{phaseop}) 
\cite{pegg}; 
and the eigenstates of the position operator 
$\hat{x}_{N} = \hat{a}_{N} + \hat{a}_{N}^{\dagger}$, for
some truncated Hilbert space ${\cal H}_{N+1}$.
The position eigenvalues are $=y$ for which
the Hermite polynomial $H_{N+1}(y) = 0$ \cite{schleich} (see Fig.\
\ref{poseig}). For large $N$ these zeros are
separated by $2\pi/\sqrt{4N}$ (asymptotic limit).
The full
Hamiltonian (\ref{Hamilton1}) was used in these simulations, and the
results
are shown in Fig.\ \ref{cat}.
It is especially interesting to consider the case of a superposition of
two
coherent states ($|\alpha\rangle + |-\alpha\rangle$). As coherent states
are
quantum-mechanical representations of essentially classical states, this
can
be considered a quantum superposition of classical states, 
or ``Schr\"{o}dinger cat'' state. Well 
separated peaks (in the position basis) can be obtained for values of
$\alpha$ as low as 1.5. The overall ``size'' of such a state is very
small 
however;
typically the peaks are separated by less than a wavelength, so that
they
cannot be observed by just looking.
Using our proposed measurement procedure, 
one would obtain in a single measurement one value of $k$ that allows us
to
elucidate one peak or the other,
as shown in Fig.\ \ref{cat}.

In summary, we have shown how to perform a single measurement for an
arbitrary
observable for the motion of a trapped ion. To our knowledge, this is
the first
example of an experimentally realistic system where arbitrary
measurements can
be practically implemented. The method is based on being able to
synthesize
arbitrary states of motion in an ion trap, and performing filtering
measurements on the state of the ion by quantum jump techniques.

We would like to thank M. Lewenstein,
R. Blatt, F. Schmidt-Kaler, T. Pellizzari and W. P.
Schleich for
discussions. This work was supported by the Austrian Science Foundation
and the
{\it Acciones Integradas} between Austria and Spain.


\begin{references}
\bibitem{degenerate} For simplicity we ignore degeneracy, but our
arguments
carry over into this case.
\bibitem{mixed} Here we use a pure state; the method also applies to
mixed
states.
\bibitem{rwa} Note that we work in  a frame rotating with the 
harmonic oscillator frequency.
\bibitem{cool} Observation of quantum jumps will heat up the ion, which
must
then be cooled to $|g,0\rangle$ before proceeding.
\end{references}
\end{document}